\documentclass[a4paper]{jpconf}
\newcommand{\hm}{H$_{2}$}
\newcommand{\apj}{ApJ}
\newcommand{\apjl}{ApJL}
\newcommand{\aap}{A\&A}

\usepackage{graphicx}
\begin{document}
\title{Molecular Hydrogen formation on grain surfaces}

\author{S. Cazaux $^1$, P. Caselli $^1$, A.G.G.M. Tielens $^2$, J. Le Bourlot $^3$, M. Walmsley $^1$}

\address{$^1$ Osservatorio Astrofisico di Arcetri, L.go E. Fermi, 5, 50125 Firenze, Italy}
\address{$^2$ Kapteyn Astronomical Institute, P.O. Box 800, 9700 AV Groningen, Netherlands}
\address{$^3$ LUTH, Observatoire de Paris, F-92195 Meudon Cedex, France}

\ead{cazaux@arcetri.astro.it}

\begin{abstract}
We reconsider \hm\ formation on grain surfaces. We develop a rate
equation model which takes into account the presence of both
physisorbed and chemisorbed sites on the surface, including quantum
mechanical tunnelling and thermal diffusion. In this study, we took
into consideration the uncertainties on the characteristics of
graphitic surfaces. We calculate the \hm\ formation efficiency with
the Langmuir Hinshelwood and Eley Rideal mechanisms, and discuss the
importance of these mechanisms for a wide range of grain and gas
temperatures. We also develop a Monte Carlo simulation to calculate
the \hm\ formation efficiency and compare the results to our rate
equation models. Our results are the following: (1) Depending on the
barrier againt chemisorption, we predict the efficiency of \hm\
formation for a wide range of grain and gas temperatures. (2) The
Eley-Rideal mechanism has an impact on the \hm\ formation efficiency
at high grain and gas temperatures. (3) The fact that we consider
chemisorption in our model makes the rate equation and Monte Carlo
approaches equivalent.
\end{abstract}

\section{Introduction}
\hm\ formation is a process which has been studied extensively in the
past decades, but which is still not well understood. In the
interstellar medium, \hm\ formation occurs on grain surfaces due to
three-body reactions which are much more efficient than gas phase
formation mechanisms (Gould \& Salpeter 1963).

Observations of \hm\ in several astrophysical environments has shown
that this molecule can be produced in various physical conditions
(Jura 1974, Hollenbach \& McKee 1979, Tielens \& Hollenbach 1985a,
1985b, Habart et al. 2004). \hm\ can form on cold and warm grain
surfaces, with various gas temperatures. Even when destroyed in
shocked regions, it will reform again in the post shocks regions. Even
in strong UV or X-ray radiation field, \hm\ may find some protected
place to form and survive. The wide variety of these environments
raises the question of how \hm\ can form in such a wide range of
physical conditions.

Theoretically, Hollenbach \& Salpeter (1970) developed a quantum
mechanical model to calculate the mobility of the atoms on a grain
surface. Because the interaction between atoms and grains involved in
their calculations were weak, they found that \hm\ formed only at low
grain temperatures. To palliate this result -- in conflict with a
variety of observations -- they took the presence of lattice defects
with enhanced binding into account (Hollenbach \& Salpeter 1971). With
this assumption, they could predict a very efficient \hm\ formation
for grain temperature $\le$ 50K.

Experimentally, Temperature Program Desorption (TPD) experiments at
low temperatures revealed the weak interactions, also called
physisorption, between the atoms and some surfaces of astrophysical
interest. Pirronello et al. (1997a, 1997b, 1999) studied the formation
of HD on olivine and carbonaceous surfaces at a range of surface
temperatures between 5 and 25K. These results allowed an estimate of
the energy of physisorption of the atoms and molecules on the surface.
Recent TPD experiments at high temperatures, performed by Zecho et
al. (2002) on graphite, showed another type of interaction between the
atoms and the surface. This interaction, also called chemisorption, is
strong and allows the formation of \hm\ at grain temperatures of
hundreds of Kelvins.

Physisorbed atoms are weakly bound to the surface and are mobile at
low grain temperatures (Ghio et al.  1980), whereas chemisorbed atoms
are strongly bound to the surface and become mobile only at grain
temperature of a few hundred K (Barlow \& Silk 1976; Aronowitch \&
Chang 1980; Klose 1992; Fromherz et al.\ 1993; Que et al. 1997;
Jeloaica \& Sidis 1999; Sha \& Jackson 2002; Cazaux \& Tielens 2002,
2004). By considering these two types of interactions between the
atoms and the surface, \hm\ can possibly form for a wide range of
temperatures.

\section{Surfaces characteristics.}

In this section, we consider carbonaceous surfaces. Olivine surfaces
have been the subject to experimental studies only at low temperatures
(Pirronello et al. 1997a, 1997b), making a characterisation of the
strong interaction between the atoms and the surface
impossible. Because of our lack of knowledge concerning chemisorption
of atoms on olivine, we will concentrate on carbonaceous
surfaces. These surfaces so called carbonaceous in amorphous form, and
graphitic in crystalline form, have been the subject of a variety of
experimental and theoretical studies (Morisset et al. 2003, Zecho et
al 2002, Sha \& Jackson 2002, Mennella et al. 1999, 2002, Pirronello
et al. 1997a, 1997b, 1999, Klose et al 1992, Aronowitch \& Chang 1980,
1985, Parneix \& Brechignac 1998). Whereas the experiments performed
at low temperatures on carbonaceous surfaces do not allow an
evaluation of the energy of chemisorption, other studies performed at
high temperatures on graphitic surfaces reveal high interactions
involved between the atoms and the surface. One of the main question
remaining is the nature of the barrier between physisorbed and
chemisorbed sites, and therefore, the barrier against
chemisorption. As shown by Zecho et al. (2002), with TPD experiments
performed at high surface and gas temperatures, the atoms from the gas
phase can, if their energy is high enough, cross the barrier against
chemisorption and occupy a chemisorbed site. Theoretically, Sha \&
Jackson (2002) and Jeloaica \& Sidis (1999), determined a high barrier
between physisorption and chemisorption of 0.2 eV. This barrier has
been determined for graphitic surfaces, and make \hm\ formation at
intermediate and high grain temperatures negligible. On the other
hand, it seems that these numerical values differ greatly depending on
the methods used. Indeed, in different theoretical studies made by
Fromhertz (1993) and Parneix \& Brechignac (1998), the barrier against
chemisorption on graphite surface is estimated to be 0.03-0.09 eV.

Finally, at elevated temperatures, H can also be chemically bonded to
a graphitic surface. The carbon atom converts then from sp$^2$ to
sp$^3$ and the hydrogen is bound by some 4 eV. Mennella et al
(2002), in a study devoted to the 3.4um aliphatic CH feature, showed
that a H/C ratio of 60 $\%$ can be attained this way.

In a previous article, we used our model to benchmark experiments at
low temperatures (Cazaux \& Tielens 2004) and characterise the weak
interactions between atoms and surface. Also, we determined a
constraint on the barrier between physisorption and chemisorption by
estimating the number of H atoms that directly chemisorbed at low
temperatures. This constraint defined only the product of the width
times the height of the barrier, but cannot separate the two (At low
temperatures, H atoms populate chemisorbed sites through tunneling
effect. This rate is a function of a$\times \sqrt(E)$, where a is the
width and E the height of the barrier between physisorbed and
chemisorbed sites). In the next sections, we therefore will consider
two extreme possibilities: (1) That described by Sha \& Jackson (2002)
and Jeloaica \& Sidis (1999) who found 0.2 eV height and suppress the
formation of \hm\ at intermediate and high temperatures. (2) That
discussed in a previous work (Cazaux \& Tielens 2004), $\sim$ 0.05 eV
height, which is consistent with the calculations of Parneix \&
Brechignac (1998) and Fromhertz (1993), and allows a high \hm\
formation efficiency, even at high gas and grain temperatures.

\section{Model for \hm\ formation: Langmuir-Hinshelwood kinetics}

We developed a rate equation model describing the formation of \hm\ on
grain surfaces. This approach is based on two main assumptions: (i)
Atom can bind to the surface in two energetically different sites: a
physisorption site (weak Van der Waals interaction) or a chemisorption
site (strong covalent bound). We assume the number of physisorbed and
chemisorbed sites on a grain to be identical. (ii) The atoms can move
from site to site by quantum tunneling or thermal diffusion.

In our model, an atom from the gas phase can accrete onto a grain only
into physisorbed sites. If the physisorbed site is already occupied,
the incoming atom bounces back to the gas phase. The accreted atoms
can scout the surface, going from site to site, and a \hm\ molecule is
formed where two atoms encounter in the same site. This process of
forming \hm\ molecules follows Langmuir-Hinshelwood kinetics. Of these
newly formed molecules, a fraction is spontaneously released in the
gas phase, and another fraction remains on the grain and evaporates if
the surface temperature is sufficiently high.

\begin{figure}
\begin{minipage}{16pc}
\includegraphics[width=16pc, angle=-90]{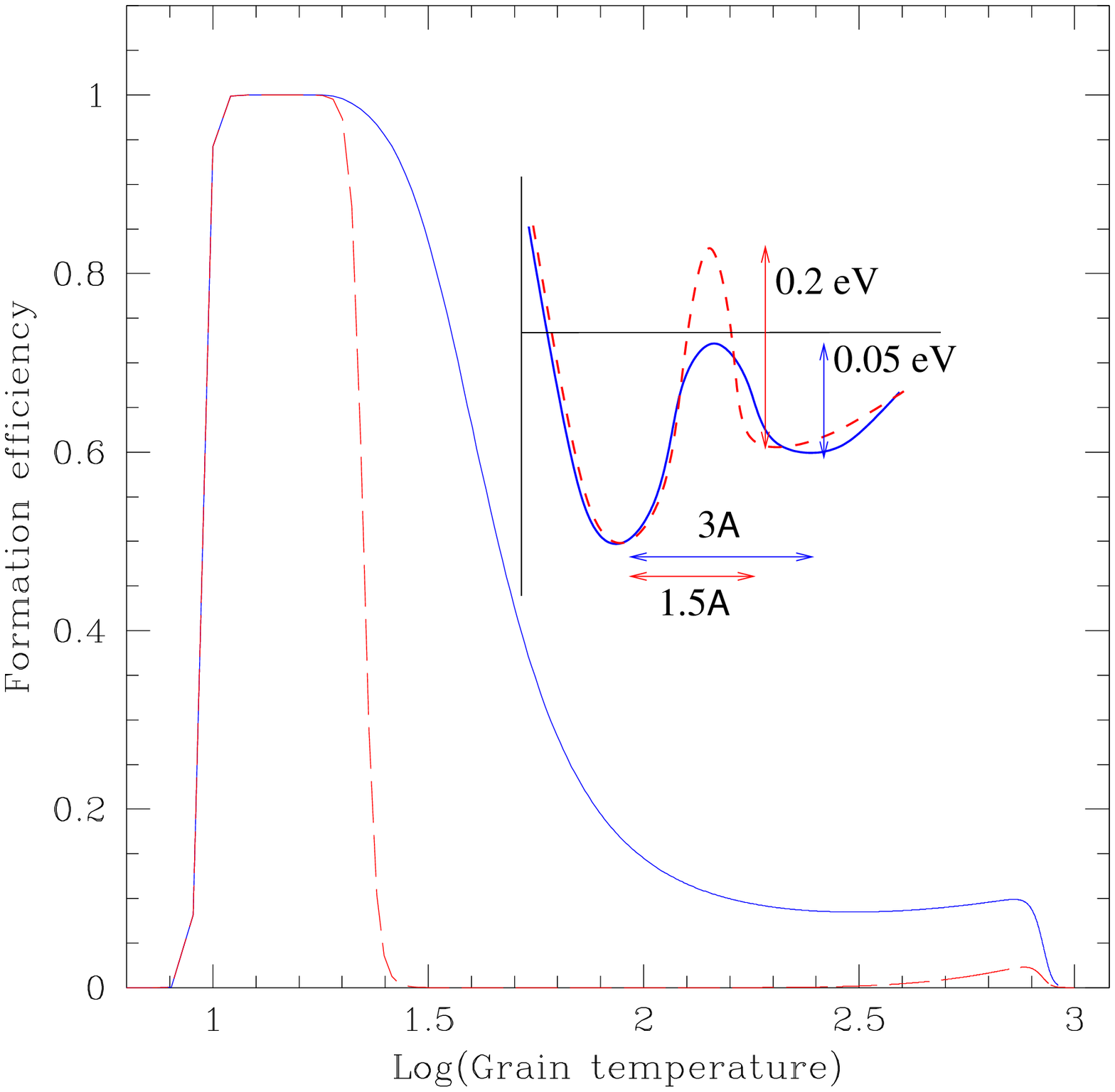}
\end{minipage}
\begin{minipage}{16pc}
\includegraphics[width=16pc, angle=-90]{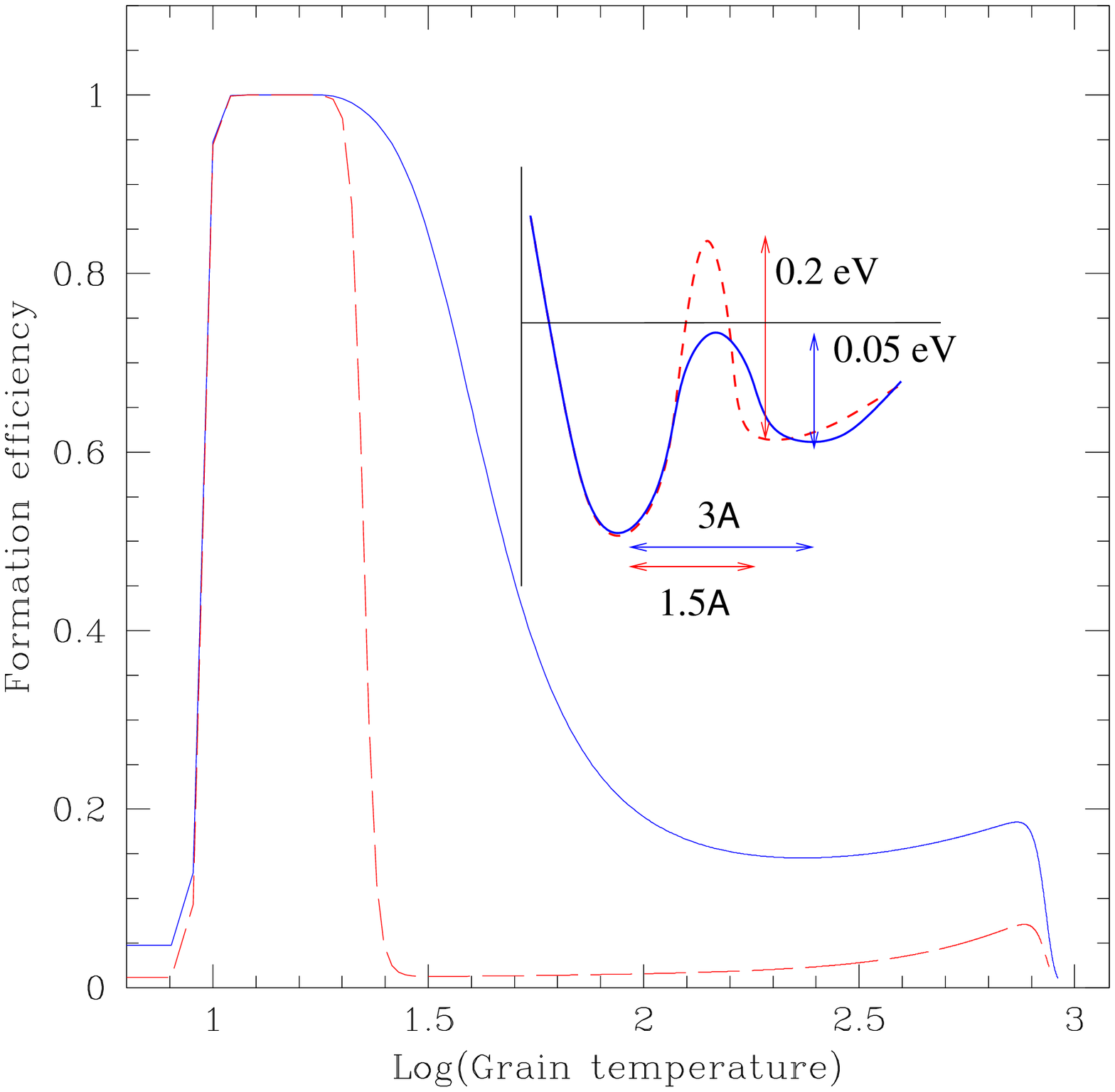}
\end{minipage}
\caption{\label{ERLK} \hm\ formation efficiency with the rate equation
model for a grain radius of 0.1 $\mu$m and a density of H atoms of
10$^2$ atoms/cm$^3$. Two extreme possible barriers between
physisorption and chemisorption are presented here, and demonstrate
that the \hm\ efficiency is either suppress if the barrier is high, or
enhance if it is low. Left: \hm\ formation efficiency considering only
Langmuir-Hinshelwood kinectics. Right: \hm\ formation efficiency
considering direct chemisorption as well as the Langmuir-Hinshelwood
and Eley Rideal mechanisms (solid lines).}
\end{figure}

We used this model to benchmark the experimental data of Pirronello et
al (1997a, 1997b, 1999). These TPD experiments probe the interactions
between the surface and the adsorbed species. Indeed, by irradiating a
surface with H atoms, and measuring the desorption of the newly formed
molecules, it is possible to characterise the interactions involved
between the atoms, the molecules and the surface. Because these
experiments were performed at very low surface temperatures (from 5 to
25 K), only the weak interactions between the atoms, the molecules and
the surface can be determined. In a previous study, we deduced the
strength of the physisorbed bounds, the nature of the barrier between
two physisorbed sites, and also gave a constraint on the barrier
between physisorbed and chemisorbed sites. Once some characteristics
of the surface under consideration are defined, it becomes possible to
calculate the \hm\ formation efficiency at steady state. In fig
~\ref{ERLK}, left panel, we report the \hm\ formation efficiency for
the two extreme barriers between physisorption and chemisorption.

The \hm\ formation efficiency at intermediate and high temperatures
remains uncertain. At very low grain temperatures, \hm\ formation
efficiency is extremely small because only a certain percentage of the
molecules formed on the grain are released in the gas phase (60 $\%$),
and the dust temperature is too low to allow evaporation of the
molecules remaining on the surface. Thus, the grain is saturated with
\hm\ molecules, suppressing adsorption of H, and becomes ``passive''
for \hm\ formation. Actually, some H atoms might bind to this ``sea of
\hm '' and lead to \hm\ formation. This aspect of \hm\ formation has
not further been investigated here. At slightly higher temperatures,
\hm\ formation efficiency rises to 100 $\%$ because the molecules
remaining on the surface can evaporate whereas the physisorbed atoms,
more strongly bound to the surface, can travel on the grain surface
until they find another H atom. Then, as temperature increases, the
physisorbed H atoms become able to evaporate before encountering
another atom. The \hm\ formation efficiency decreases, and \hm\ forms
through the association of physisorbed and chemisorbed atoms, at
intermediate temperatures, and then through association of chemisorbed
atoms at high temperatures.

\section{Langmuir Hinshelwood VS Eley Rideal}
In this section, we consider another possible way to form \hm\
molecules on grain surfaces: the Eley Rideal mechanism which is
relevant at high grain temperatures (Parneix \& Brechignac 1998, Zecho
et al. 2002) and when the collision energy of the incoming atom is
high enough (Morrisset et al. 2003). In this approach, the \hm\
molecules can be formed if an atom coming from the gas phase arrives
in an already occupied site. Also, we consider possible direct
chemisorption: if an incoming atom possesses enough energy to directly
cross the barrier against chemisorption, it can directly be
chemisorbed. Both mechanisms increase the \hm\ formation efficiency at
high temperatures.


In fig ~\ref{ERLK}, right panel, we report the \hm\ formation
efficiency considering direct chemisorption as well as
Langmuir-Hinshelwood and Eley-Rideal mechanisms, when the rate
equation system reach a steady state equilibrium. The importance of
the barrier between physisorbed and chemisorbed sites is striking. If
this barrier is high (0.2 eV height and 1.5 \AA\ width), the \hm\
formation efficiency is almost suppressed, even considering both
Langmuir-Hinshelwood and Eley Rideal mechanisms. If, on the contrary,
the barrier is low (0.05 eV and 3 \AA\ width), the \hm\ formation
efficiency is enhanced considerably. In our calculations, we consider
the gas and grain temperatures coupled. Obviously, an increased of the
gas temperature will enhance the Eley Rideal mechanism as well as the
direct chemisorption.

\begin{figure}
\begin{minipage}{16pc}
\includegraphics[width=16pc, angle=-90]{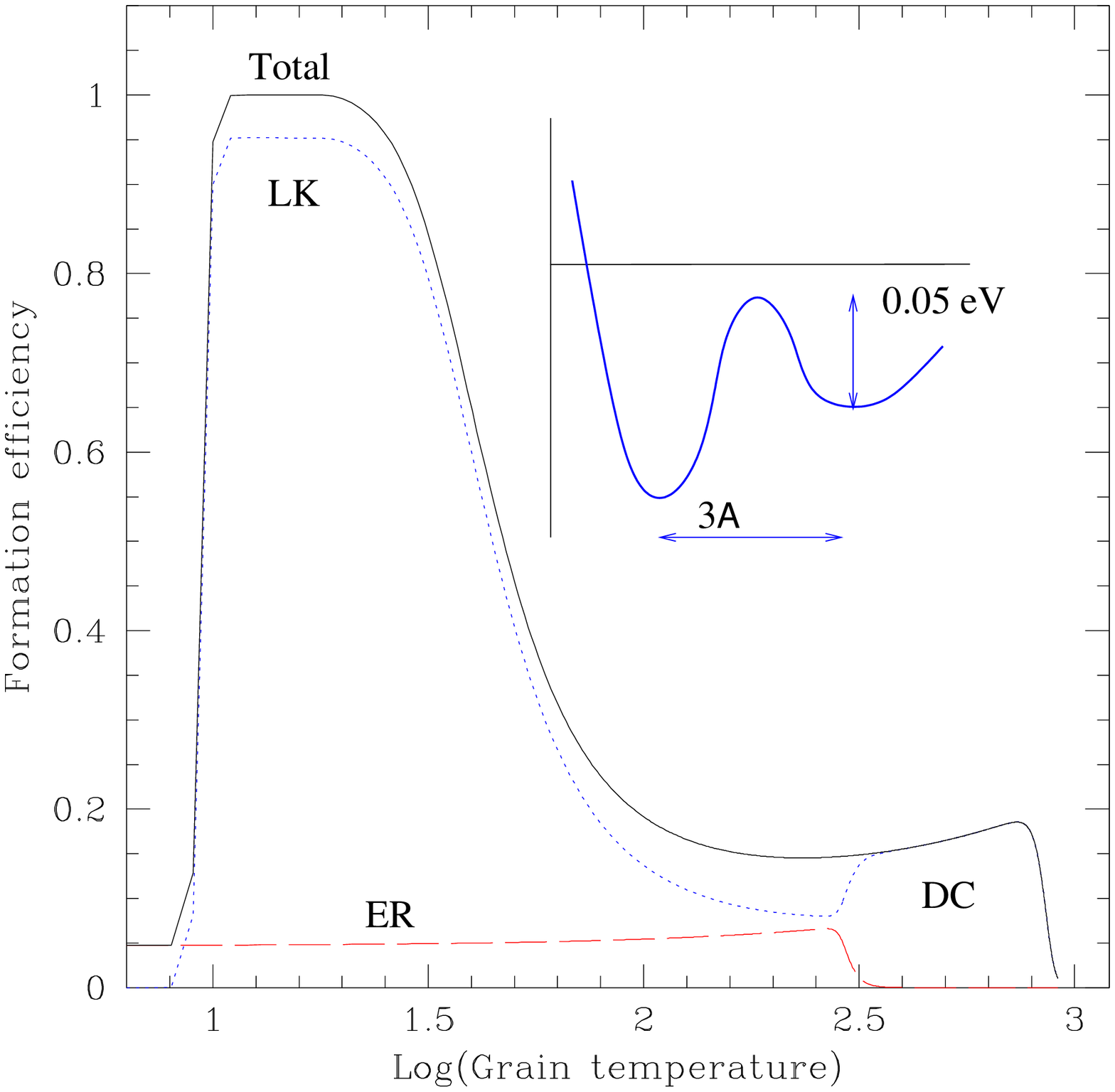}
\end{minipage}
\begin{minipage}{16pc}
\includegraphics[width=16pc, angle=-90]{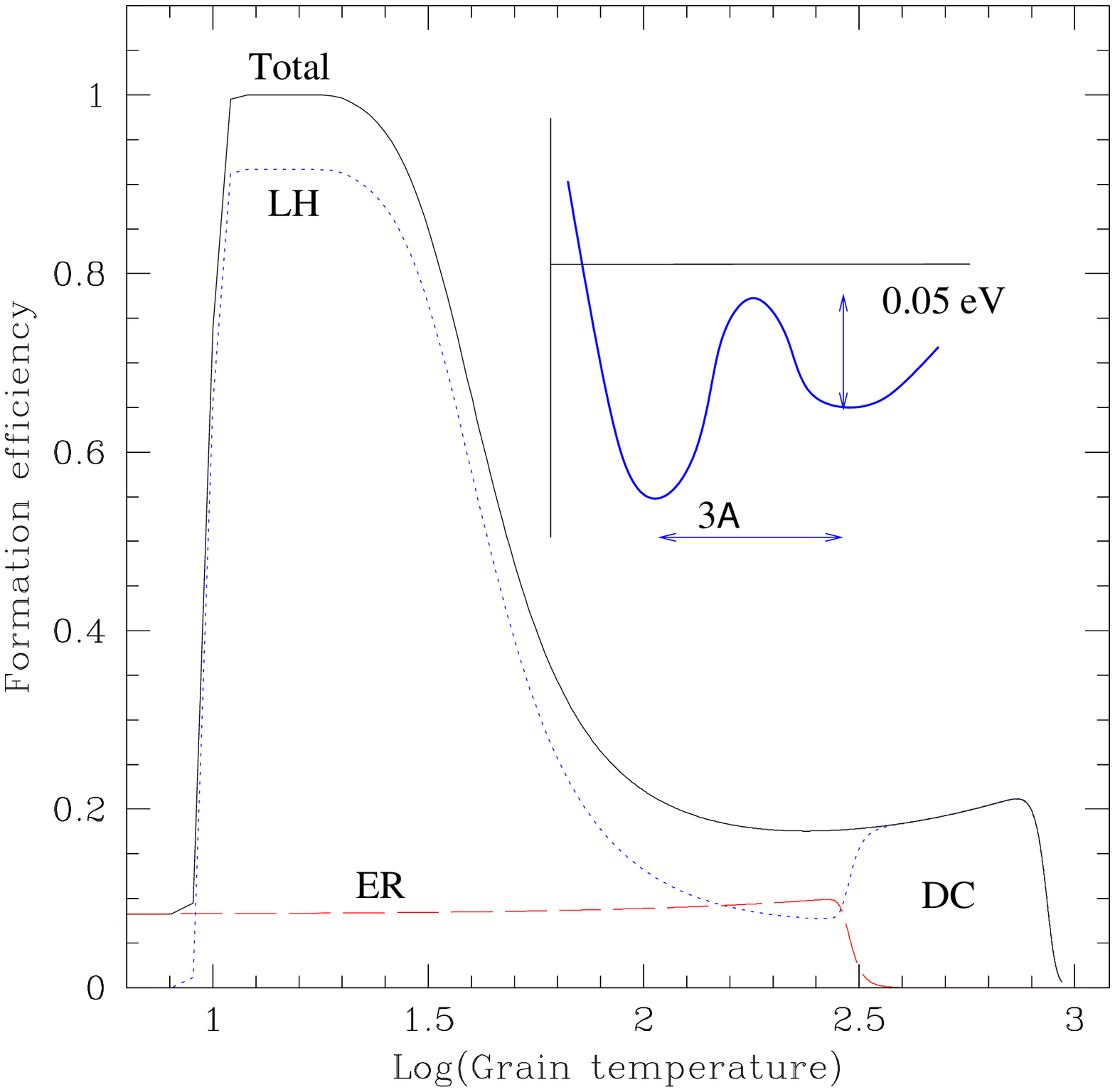 }
\end{minipage}
\caption{\label{ERVSLK} \hm\ formation efficiency due to
Langmuir-Hinshelwood (LH) and Eley-Rideal (ER) mechanisms are
reprensented. The direct chemisorption (DC) mechanism populates the
chemisorbed sites and increases the efficiency of the
Langmuir-Hinshelwood mechanism at high grain temperature. The
temperature of the gas and the grains are set as: Left: T$_{gas}$ =
T$_{dust}$. Right: T$_{gas}$ = T$_{dust}$ + 500K.}
\end{figure}

Fig ~\ref{ERVSLK} shows at which temperatures which mechanisms are
predominant for \hm\ formation. It is clear that the Eley Rideal
mechanism is important at low and intermediate temperatures. This
mechanism stops when the chemisorbed atoms become mobile enough to
scout the grain surface and associate with another H atom. The
Langmuir-Hinshelwood mechanism is predominant in the \hm\
formation. At high temperatures, the H atoms populate the chemisorbed
sites by direct chemisorption, scout the surface of the grain, and
associate through Langmuir-Hinshelwood mechanism. The gas temperature,
as shown in fig ~\ref{ERVSLK} left and right panels, changes slightly
the efficiency of the different mechanims.

\section{Monte Carlo simulation}
\begin{figure}
\begin{minipage}{20pc}
\includegraphics[width=14pc, angle=0]{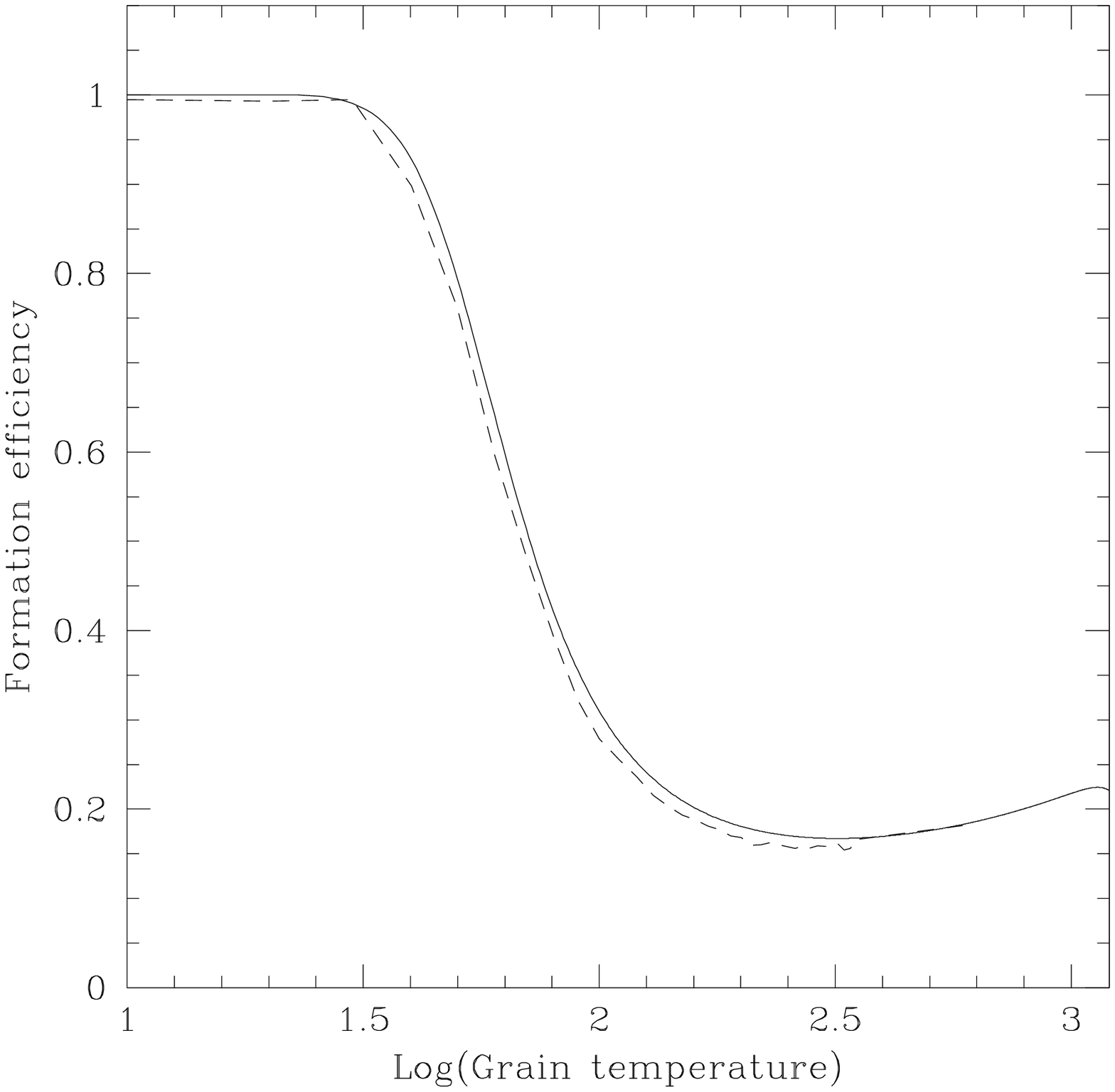}
\end{minipage}
\begin{minipage}{20pc}
\includegraphics[width=14pc, angle=0]{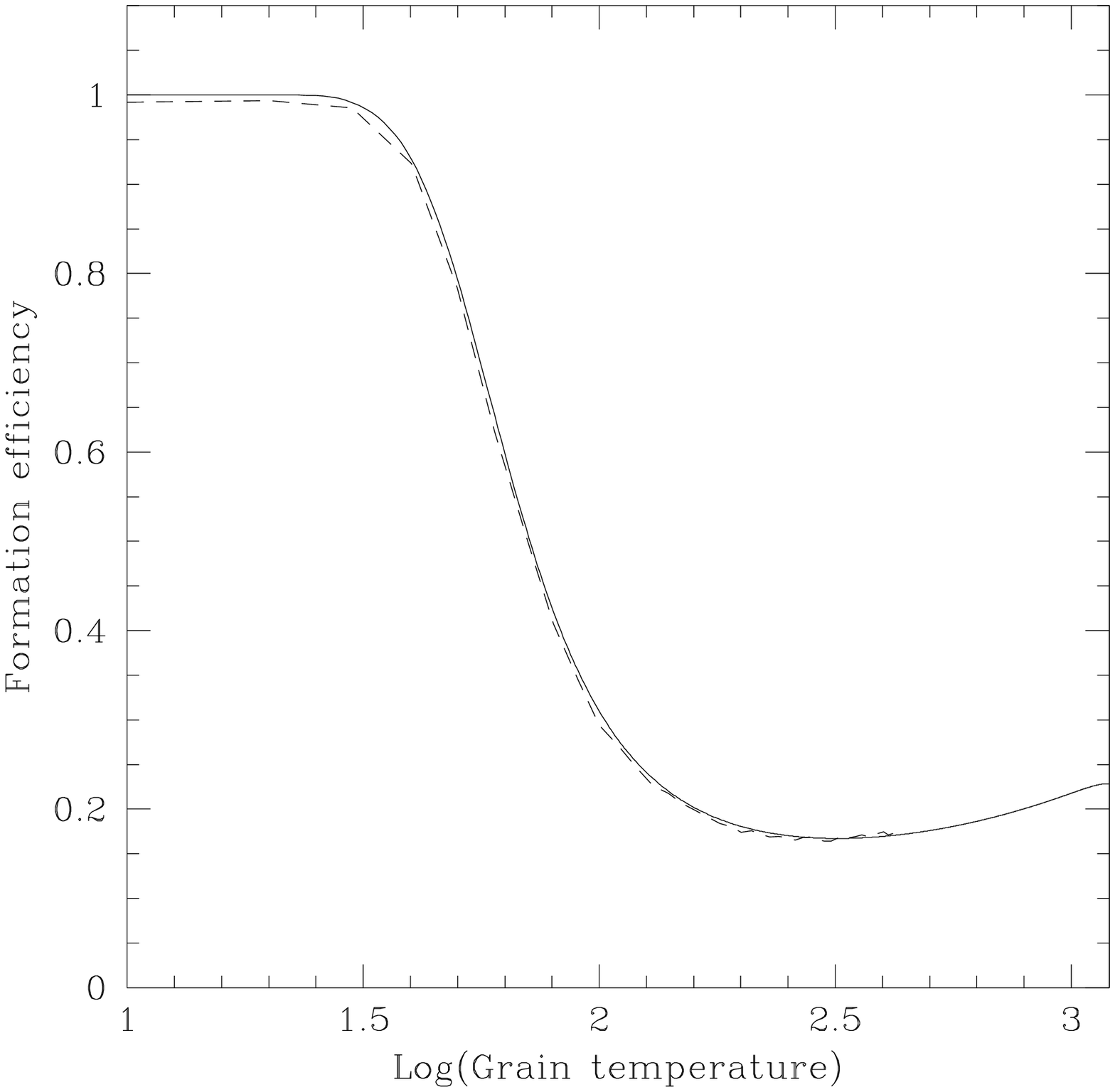}
\end{minipage}
\caption{\label{compa}Comparison of the \hm\
formation efficiency in steady state equilibrium obtained with a rate
equations (solid line) and a Monte Carlo (dashed lines)
approachs. These efficiencies are calculated for a grain with a radius
of 100\AA (left panel), and with a radius of 20 \AA (right panel).}
\end{figure}

We have developed a Monte Carlo simulation to describe \hm\ formation
on grain surfaces. The grain is seen as a squared grid, with, at each
of its points, the possibility to have a chemisorbed and physisorbed
atom. We consider as before direct chemisorption as well as
Langmuir-Hinshelwood and Eley-Rideal mechanisms. When an atom comes
from the gas phase onto this grid, depending on its energy, it can
become physisorbed or chemisorbed. If the site is already occupied, it
can form a molecule which is released into the gas phase. The position
of the incoming atom on the grid is chosen randomly. Once on the grid,
the atom can move from site to site according to its energy and
follows a random walk. If two atoms arrive in the same site, they
associate to form a molecule. As we did previously with the rate
equation model, we report the \hm\ formation efficiency when the Monte
Carlo model reach the steady state equilibrium. In fig ~\ref{compa},
we consider grains with a radius of 100 \AA\ (left panel) and with a
radius of 20 \AA\ (right panel).  We compare the \hm\ formation
efficiency in steady state equilibrium obtained by the rate equations
and Monte Carlo methods. The two methods give identical results until
quite high temperatures ($\sim$400K). At very high temperatures we
expect the two models to differ because we arrive in a regime where
the grain has 1 atom or less. As a conclusion, for a large range of
physical conditions, the presence of chemisorbed sites on the grain
surface insures the coverage to be high enough that the Monte Carlo and
rate equations approachs are equivalent.

\section{Conclusions}
We reconsidered the model of \hm\ formation developed in Cazaux \&
Tielens (2004). We discussed the uncertainties on the characteristics
of grain surfaces and the nature of the barrier between physisorbed
and chemisorbed sites. A barrier which is too high prevents the atoms
entering into chemisorbed sites, and therefore supresses \hm\
formation at high grain temperatures. A low barrier, on the contrary,
allows the atoms to populate these sites, and enhances \hm\ formation
at high temperatures through the recombination of chemisorbed
atoms. We also compared two mechanisms of \hm\ formation, the
Langmuir-Hinshelwood and Eley Rideal processes. The Eley Rideal
mechanism gives a small contribution to the \hm\ formation efficiency
at low and intermediate temperatures (T$_{grain}$ $\le$ 300K ). At high
temperatures, the chemisorbed atoms become mobile and
Langmuir-Hinshelwood mechanism dominates completly. We developed a
Monte Carlo simulation to calculate \hm\ formation efficiency on small
grains. These results are in perfect agreement with the rate equations
model until the regime where the grain possess 1 or less atom is
reached. Thus, the inclusion of chemisorption in our model makes the
\hm\ formation efficiency through rate equations or Monte Carlo
simulations equivalent.

\end{document}